\documentclass[epj,nopacs]{svjour}
\pdfoutput=1
\usepackage[utf8]{inputenc}
\usepackage{caption,amsmath,amssymb,amsfonts,graphicx,cite,multirow,tikz,enumitem}
\DeclareCaptionType{Diagram}
\usepackage{pstricks,pdflscape,morefloats,algorithmicx,algpseudocode,slashed,todonotes}
\usepackage[section]{algorithm}
\usepackage{booktabs}

\usepackage{color}
\usepackage{hyperref}

\usetikzlibrary{trees}
\usetikzlibrary{decorations.pathmorphing}
\usetikzlibrary{decorations.markings}

\usepackage{enumitem,mathtools}

\makeatletter
\def\cl@chapter{\@elt {theorem}}
\makeatother

\newcommand{\professor}{\texttt{Professor}}
\newcommand{\rivet}{\texttt{Rivet}}

\usepackage{subfigure}
\usepackage{graphicx} 

\usepackage[switch,pagewise]{lineno}

\preprint{LU-TP-19-53, KA-TP-23-2019, MCNET-19-25, HERWIG-2019-2}

\title{Improving the description of multiple interactions in Herwig}

\author{
Johannes Bellm\inst{1}\and 
Stefan Gieseke\inst{2} \and 
Patrick Kirchgaesser \inst{2}
}

\institute{
Theoretical Particle Physics, Department of Astronomy and Theoretical Physics, Lund University, Lund, Sweden  \and Institute for Theoretical Physics, Karlsruhe Institute of Technology, 76128 Karlsruhe, Germany
}

\date{\today}

\abstract{
The modelling of multiple parton interactions in Monte Carlo event generators is a crucial part not only 
for the dressing of signal processes but also to describe data with a minimum bias on the event selection. 
Much work has and will be put into the theoretical framework and the numerical implementation of these models.
In this contribution, we summarize various changes to the machinery of multiparton 
interactions and related physics in the Monte Carlo event generator Herwig 7 \cite{Bellm:2015jjp}.
}

\begin{document}

\maketitle

\section{Introduction and Motivation} 
\label{sec:intro}

Due to the composite nature of hadrons, it is possible to have multiple parton interactions (MPI) during a single scattering event. 
The nature of these interactions
is responsible for the various final state topologies which are being measured at hadronic collisions. Final states with an accumulation of several particles with high transverse momentum ($p_{\perp}$) are typically referred to as \textit{jets} and can be described reliably with the methods of perturbative QCD. 
On the other hand, when no high $p_{\perp}$ particles are present, the event is usually characterized by large hadronic activity, distributed flat in rapidity with a relatively low $p_{\perp}$. These events are attributed to the soft regime
where the methods of perturbative QCD break down and one has to rely on modelling. 
This distinction in hard and soft events is of course not always clear. Especially in the
transition region between hard and soft interactions, the correct interplay of the models becomes crucial. For a full description of the experimental data available, diffractive events need to be incorporated into the simulation chain as well. 
Multi purpose event generators like Herwig \cite{Bellm:2015jjp,Bellm:2017bvx}, Sherpa \cite{Bothmann:2019yzt} and Pythia\cite{Sjostrand:2006za,Sjostrand:2014zea}, all have models that model 
this part of the hadronic interactions \cite{Sjostrand:1985vv,Sjostrand:1987su,Sjostrand:2004pf,Bahr:2008wk,Bahr:2008dy,Corke:2009tk,Corke:2011yy,Martin:2012nm}.
The model for hard MPI was introduced in the newer C++ Version of
Herwig \,in \cite{Bahr:2008wk,Bahr:2008dy} and extended to the soft regime in 
\cite{Bahr:2009ek}. The additional soft and hard interactions are seamlessly
integrated into the existing framework and lie at the intersection
between the parton shower of the hard process and the hadronization
model. The soft interactions have been revised in
\cite{Gieseke:2016fpz} and replaced with a particle ladder obeying multiperipheral
kinematics \cite{Amati:1962jaa,Baker:1976cv}. 
The combination with the newly introduced model for diffraction led
to an improved description of the majority of Minimum Bias (MB) and Underlying Event (UE)
 observables \cite{Gieseke:2016fpz}.
Recently and already including various of the changes described in this contribution, 
information on the space-time structure of MPIs and colour reconnection 
was introduced in \cite{Bellm:2019wrh}.
In this paper, we attempt to summarize the various changes to the MPI
in Herwig \cite{Bellm:2015jjp,Bellm:2017bvx}  
that have accumulated over the past months and resulted in an 
improved description of various UE and MB observables.
The paper is structured as follows. In Sec.~\ref{sec:theproblem} we review the current state
of the multi parton interaction model in Herwig, with a focus on the event generation
workflow and the technicalities of the simulation.
In Sec.~\ref{sec:modifications} we summarize the modifications to the MPI model.
After describing the tuning procedure in Sec~\ref{sec:tuning}, we show the impact of the modifications
and compare to data in Sec.~\ref{sec:results}.

\section{Current State}
\label{sec:theproblem}

\tikzset{
particle/.style={thick,draw=blue, postaction={decorate},
    decoration={markings,mark=at position .5 with {\arrow[blue]{triangle 45}}}},
gluon/.style={decorate, draw=black,
    decoration={coil,aspect=0}}
 }
\tikzset{
particle/.style={thin,draw=blue, postaction={decorate},
decoration={markings,mark=at position .5 with {\arrow[blue]{stealth}}}},
gluon/.style={decorate, draw=black,
    decoration={coil,amplitude=4pt, segment length=5pt}}
}
 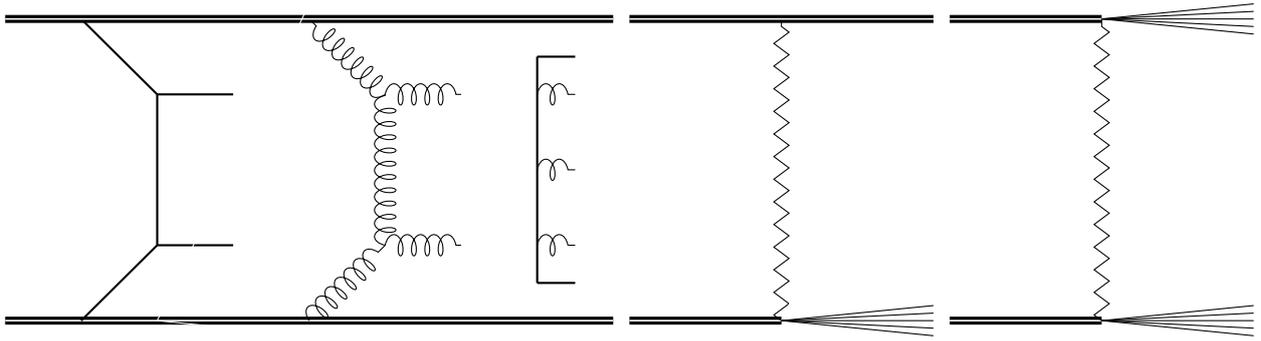
\begin{figure*}[t]
  \centering
\begin{tikzpicture}
\draw [double,very thick](0,0) -- (8,0);
\draw [thick](1,0) -- (2,1);
\draw [thick](2,1) -- (2,3);
\draw [thick](2,3) -- (1,4);
\draw [thick](2,3) -- (3,3);
\draw [thick](2,1) -- (3,1);
\draw [gluon](4,0) -- (5,1);
\draw [gluon](5,1) -- (5,3);
\draw [gluon](5,3) -- (4,4);
\draw [gluon](5,3) -- (6,3);
\draw [gluon](5,1) -- (6,1);
\draw [thick](7,0.5) -- (7,3.5);
\draw [thick](7.,0.5) -- (7.5,0.5);
\draw [thick](7.,3.5) -- (7.5,3.5);
\draw [gluon](7.,1.) -- (7.5,1.);
\draw [gluon](7.,2.) -- (7.5,2.);
\draw [gluon](7.,3.) -- (7.5,3.);
\draw [double,very thick](0,4) -- (8,4);      
\draw[white] (2,0) -- (4,-0.2);
\draw[white] (2,0) -- (4,4.2);
\end{tikzpicture}
\begin{tikzpicture}
\draw [double,very thick](0,0) -- (2,0);

\draw [decoration = 
{zigzag,segment length = 3mm, amplitude = 1mm},decorate](2,0) -- (2,4);
\draw[white] (2,0) -- (4,4.2);
\draw (2,0) -- (4,-0.2);
\draw (2,0) -- (4,-0.1);
\draw (2,0) -- (4,0);
\draw (2,0) -- (4,0.1);
\draw (2,0) -- (4,0.2);
\draw [double,very thick](0,4) -- (4,4);      
\end{tikzpicture}
\begin{tikzpicture}
\draw [double,very thick](0,0) -- (2,0);
\draw [decoration = 
{zigzag,segment length = 3mm, amplitude = 1mm},decorate](2,0) -- (2,4);

\draw (2,0) -- (4,-0.2);
\draw (2,0) -- (4,-0.1);
\draw (2,0) -- (4,0);
\draw (2,0) -- (4,0.1);
\draw (2,0) -- (4,0.2);
\draw [double,very thick](0,4) -- (2,4); 

\draw (2,4) -- (4,4.2);
\draw (2,4) -- (4,4.1);
\draw (2,4) -- (4,4);
\draw (2,4) -- (4,3.9);
\draw (2,4) -- (4,3.8);
     
\end{tikzpicture}
\caption{Pictorial representation of the processes contributing to the modelling 
of minimum bias production in  Herwig. On the left-hand side the 'dummy' process to start hard (QCD two-to-two process) and soft (quark pair with gluon ladder) multiple parton interactions. On the right the two possible single and double diffraction processes. If no hard scatter is produced the incoming beams are modelled without colour exchange. }
\label{fig:picmpi}
\end{figure*}

For a detailed description of the theoretical basis of the MPI model, 
we refer to \cite{Bahr:2008pv}. In the following, we give an overview of the different building blocks of the simulation in order to put multi parton interactions in the right context. 
As it stands the MPI framework in Herwig is split into two parts: 
First, the non-diffractive part (ND), with a hard interaction that breaks up both protons of the incoming state and requires at least one additional scatter. 
Second, the diffractive part that keeps either one of the incoming protons intact 
(Single diffraction (SD)) or breaks both protons (double-diffraction (DD)) without colour exchange. 
In the simulation of ND events with Herwig, the starting
point is a dummy process (MEMinBias)
that was introduced to split up the protons and replace the hard process. 
Here the process extracts (anti-)quarks from the protons without transverse momentum or parton showering. As by construction, the hard process in Herwig is forced to evolve back to a valence quark, the possible forced splittings are performed based on probabilities calculated from DGLAP splitting kernels. 
Once this is performed the number of additional soft and hard scatters is determined \cite{Bahr:2008pv}. 
 The additional hard scatters constructed from regular QCD two-to-two processes are --
  including shower and backward evolution to gluons --  attached to the beam remnant. 
If the showering process or possible forced splitting to gluons creates states that are not compatible with the already extracted energy fraction, the additional scatter is discarded and a new trial is performed. 
Once the hard additional scatters are added, the RemnantHandler of Herwig adds the additional soft scatters to the process. It is possible to
choose between the traditional soft two-to-two processes or the multiperipheral model introduced in \cite{Gieseke:2016fpz}. 
We will concentrate on the model described in \cite{Gieseke:2016fpz} as it solved long-standing issues in the description of rapidity gap observables reported in \cite{Aad:2012pw}.
The multiperipheral model \cite{Gieseke:2016fpz} makes use of the same transverse momentum distribution as the traditional two-to-two soft model but creates gluon ladders that are distributed equidistantly in rapidity placed between a quark-antiquark pair. 
 At this point, the ladders are located as color singlets randomly between the current beam remnants. To ensure momentum conservation the simplistic choice has been made to correlate the gluons pairwise. As a consequence of this unphysical correlation observables can be constructed to identify this feature. 
 Once the additional soft and hard scatters are attached to the event and beam remnant 
 the hadronization and color reconnection models take over.
 The second component, namely the diffractive events are constructed using the 
 MEDiffraction Matrix element of Herwig. A detailed description of kinematics and particle production is given in \cite{Gieseke:2016fpz}. For the understanding of the following paragraphs, only the overall normalization of the diffractive cross-section is of relevance as hardly any changes have been made to the modeling. In the current state, 
 the MEMinBias is weighted with a parameter named CSNorm and the diffractive 
 cross-section comes with
   additional factors that can be modified. To ensure the overall cross-section a 
   cross-section reweigther was introduced to ensure that the sum of SD, DD, and ND 
   add up to the inelastic cross-section. 
   
 To illustrate the various components of the MinBias production mechanisms  Fig.~\ref{fig:picmpi} shows from left to right the 
 dummy process that comes with hard (QCD two-to-two like) and soft (quark pair with gluon-ladder) non-diffractive processes as well as the SD and DD processes. 

\section{Modifications}
\label{sec:modifications}

In this section, we describe modifications to the existing algorithms, 
starting with changes that are supposed to keep the result unmodified but improve the ability to produce events. 
We then focus on changes that will modify the physics in a motivated manner.

\textbf{Reweighting:} 
Before Herwig 7.2, a PostProcess-\\Reweighter was introduced to modify the event weights such that the cross-section calculated in the MPIHandler restores the inelastic cross-section.
In this construction, events with variable weights, that can even be negative, were produced.

To circumvent this behaviour, we modified the matrix elements. They now keep track of their cross-section and reweight the next events such that the desired cross-section is produced on the hard cross-section level. The net effect is that the reweighting is pulled to the stage before unweighting such that the produced and further evolved events have unit weights. Even though the change seems trivial, the result improves the performance of event generation and tuning. 

For the reweighting we solve,
\begin{equation}
  (N+\Delta )\cdot \sigma_{D} = N\cdot \sigma_{C}  + w\cdot \Delta\cdot  \sigma_{\mathrm{no rew.}} \nonumber
\end{equation}
for $w$, where $\sigma_{D}$ is the desired,  $\sigma_{C} $ is the current cross-section and the pair $N$ and $\Delta$ are the current number of points and the weight updating interval respectively, we choose $\Delta=50$.
The cross-section without reweigthing $\sigma_{\mathrm{no rew.}}$ is, for large number of points $N$,
\begin{equation}
\sigma_{C} = \bar{w}\cdot \sigma_{\mathrm{no rew.}}\;,
\end{equation}
where $\bar{w} $ is the average weight calculated as $\sum w / N$.

\textbf{DiffractionRatio:}
To vary the ratio between non-diffractive and diffractive events the parameter CSNorm in the matrix element to produce the non-diffractive cross-section was used. 
We change this behaviour and introduce a 
parameter that gives the fraction of diffractive cross-section with respect to the non-diffractive cross-section, 
named the DiffractionRatio. As the cross-section of the dummy process was constructed such that scale and cuts on the incoming momenta fraction can influence the unphysical -- that is reweighted to the physical -- cross-section it is beneficial to parametrize the influential quantity, to begin with. 
 
The cross-section of hard and soft MPIs that are used in the eikonal model is now adjusted such that the sum, after eikonalisation gives the total cross-section. As in this model, the diffractive part is seen to be included in the 
two components, the DiffractionRatio now takes out a part of the non-elastic cross-section after eikonalisation.

\textbf{Partner and Scale Choice:}
The shower starting conditions are a delicate problem and especially in the case of hard MPIs it is possible to enhance or reduce the amount of additional radiation. 
Further, the recoil of possible emissions and the kinematic reconstruction can be modified in various schemes that will not modify the accuracy of the showering process. 
In earlier versions, it was argued that the color partner of a radiating parton should be chosen such that the angle is maximal and the scale of the emitter is chosen with respect to the partner parton. 
We modify this to a scheme that chooses the evolution partner of the gluons randomly and then, as before, choosing the scale with respect to this chosen partner. 
This scheme is already the standard when using NLO corrections or external LHE-files for event generation in Herwig. 
In Sec.~\ref{sec:results} we also discuss the possibility of choosing the scale different from the evolution partner.

\textbf{Dummy Process using Valence Quarks:}
In the Herwig event generation for the description of minimum bias data a hard dummy process 
was introduced to keep the default workflow and enable event generation with a hard sub-process.
The matrix element that was introduced to hardly alter the observed final state returns parton configurations that have zero transverse momentum and an energy fraction that is basically given by the parton distribution functions. 
We observed that the number of trials needed to generate the average number of hard MPIs $\langle n_{\mathrm{hard}}\rangle$ 
strongly exceeds $\langle n_{\mathrm{hard}} \rangle$. The reason for this can be found in the dummy events that are produced with sea quark content. Those events are, as the Herwig remnant will expect the primary hard process to end on a valence quark, forced to split back to a valence quark.
As the leg(s) with sea-quark content will not perform a showering (zero $p_t$), 
it will need two forced splittings -- first to a gluon and then to a valence quark -- in order to be extracted from the remnant. This procedure takes a large portion of the energy of the proton, such that additional scatters, if they are required to be added, will be forced/biased to have smaller
momentum fractions and therefore produce a softer final state. 
Requiring the (anti-)quarks to match the valence quark content of the (anti-)proton
will modify the spectrum as can be seen in Fig.~\ref{fig:valence}.

\textbf{Kinematics of Soft Ladder:}
The particle kinematics in the soft ladder were generated following the algorithm presented in \cite{Baker:1976cv}. This approach led to highly anticorrelated mini-jet events as was pointed out in \cite{Azarkin:2018cmr}. 
This issue has been resolved with an improved algorithm based on the ideas in \cite{Jadach:1975woe}. We have implemented this model after initial studies in \cite{Gay:Thesis:2018} 
The number $N$ of the particles in the soft ladder is drawn from a Poissonian distribution
with mean
\begin{equation}
\langle N \rangle = N_{\mathrm{ladder}}\ln \frac{(p_{r1}+p_{r2})^2}{m_{\mathrm{rem}^2}} + B_{\mathrm{ladder}},
\label{eq:ladderMult}
\end{equation}
where $N_{\mathrm{ladder}}$ and $B_{\mathrm{ladder}}$ are parameters which will be tuned
to MB and UE data, $p_{r1,2}$ are the momenta of the incoming remnants and $m_{\mathrm{rem}}$
is the constituent mass of the remnant.
Instead of directly calculating the momentum fraction necessary to
distribute the particles equally spaced in rapidity, we sample the rapidities of
the particles flat in the available rapidity interval $-Y_{\mathrm{max}}<y<Y_{\mathrm{max}}$ defined by the two beam remnants. This enhances the variance in the rapidity compared to the
previous algorithm where the rapidity values essentially were pre-determined by the available
energy in the beam remnant system. 

\textbf{Soft Ladder Transverse Momentum:}
Another significant change to the model for soft interactions affects the transverse momentum
distribution of the ladder particles. 
Instead of drawing the $p_{\perp}$ of every particle
in the ladder from the distribution 
\begin{equation}
\frac{\mathrm{d}\sigma_{\mathrm{soft}}}{\mathrm{d}p_{\perp}^2}\sim 
p_{\perp}e^{-\beta(p_{\perp}^2-p_{\perp}^{\mathrm{min,2}})},
\label{eq:pTmin}
\end{equation}
only one $p_{\perp}$ of the soft ladder particles is sampled 
according to Eq.\,~\ref{eq:pTmin}. The $p_{\perp}$ of the remaining
particles have to fulfill the requirement that $p_{\perp,2..n}<p_{\perp,1}$.
This is necessary in order not to bias the resulting 
$p_{\perp}$ distribution towards higher values and to 
reproduce the shape of the $\frac{d \sigma}{d p_{\perp}^2}$ 
distribution as calculated within the eikonal model (see \cite{Bahr:2008pv}).
This change in assigning the $p_{\perp}$ values to the ladder 
particles leads to a significant improvement in the 
low multiplicity region of the 
$\langle p_{\perp} \rangle\, \mathrm{vs.}\,N_{\mathrm{ch}}$ observable 
as is shown in Fig.\,~\ref{fig:avgPtNch}.

\textbf{Modified Power Law for $p_{\perp}^{\mathrm{min}}$(s):}
The power law for the energy extrapolation of the parameter $p_{\perp}^{\mathrm{min}}$(s) was found to describe data at high centre-of-mass energies.
Incorporating lower centre-of-mass energies proved to be difficult since with
the existing power-law, the eikonal model could not be solved in a consistent way.
A dedicated tuning procedure led to 
a modified power-law including $p_{\perp}^{\mathrm{min}}$(s) values tuned to MB and
UE data from centre-of-mass energies below 900 GeV. 
The modified power law for the $p_{\perp}^{\mathrm{min}}$ parametrization now reads
\begin{equation} 
\label{eq:powerlaw}
p_{\perp}^{\mathrm{min}}(s) = p_{\perp,0}^{\mathrm{min}} \left( \frac{b+\sqrt{s}}{E_0} \right)^c,
\end{equation}
where $b$ is the offset necessary to fit the $p_{\perp}^{\mathrm{min}}$
values for small $\sqrt{s}$. 
A detailed description of the tuning procedure and the 
parametrization is outlined in Sec.\,~\ref{sec:tuning}.

\textbf{Colour Disrupt:}
Furthermore, several attempts were undertaken to incorporate different colour
connection topologies similar to \cite{Gieseke:2016pbi} in order to minimize the number
of MPI events containing sizeable rapidity gaps $\Delta \eta$, mimicking diffractive processes.
But none of the tested topologies involving a colour connection between the beam 
remnant and the particle ladder resulted in an improved description of data.
We, therefore, stick to the current implementation which treats the particle ladder
as a colour singlet and connects the beam remnant to the hard part of the event\footnote{Correlations in the final state are introduced via colour reconnection.}.

\section{Tuning to MB and UE data}
\label{sec:tuning}
Since there were several changes to the underlying structure of the MPI model and the soft part of the MPI model itself, a retune of the model parameters is necessary. We tune the model to MB and UE data covering the centre-of-mass energy, 
$\sqrt{s}$, between 200\,GeV and 13\,TeV. The tuning was performed by using the 
\rivet\,\cite{Buckley:2010ar} and \professor\,\cite{Buckley:2009bj} frameworks 
for Monte-Carlo event generators in combination with Autotunes\,\cite{Bellm:2019owc}.
The parameters considered in the tuning are the main parameters of the MPI model, the 
minimum transverse momentum $p_{\perp}^{\mathrm{min}}$ and the inverse proton radius 
squared $\mu^2$, the parameters determining the number of partons in the soft ladder, 
$N_{\mathrm{ladder}}$ and $B_{\mathrm{ladder}}$ (see Eq.\,~\ref{eq:ladderMult}), 
the fraction of the diffractive cross-section from the inelastic cross-section 
$R_{\mathrm{Diffraction}}$ and the two colour reconnection probabilities 
$p_{\mathrm{Reco}}$, $p_{\mathrm{RecoBaryonic}}$ of the modified colour reconnection model which was introduced in \cite{Gieseke:2017clv}. 
We start the tuning with a reassessment of the energy dependence of the 
minimum transverse momentum $p_{\perp}^{\mathrm{min}}$ and the inverse proton 
radius squared $\mu^2$. The two parameters are the main parameters of the MPI model and 
are responsible for the interplay between hard and soft interactions by directly stirring the
relevant cross-sections.
We tune both values to MB and UE data from 200\,GeV\,\cite{Abelev:2008ab,Albajar:1989an}, 500\,GeV\,\cite{Albajar:1989an}, 900\,GeV\,\cite{Aad:2010fh,Aad:2010ac,Aad:2011qe,Aamodt:2010pp}, 1.8\,TeV\,\cite{Abe:1988yu,Abe:1989td,Acosta:2001rm,Affolder:2001xt,Acosta:2004wqa}, 7\,TeV\,\cite{Adam:2015qaa,Aamodt:2010pp,Aad:2011qe,Aad:2010ac,Aad:2010fh,Aad:2012pw,Khachatryan:2011tm} and 
13\,TeV \,\cite{Aaboud:2017fwp,Aaboud:2016itf,Sirunyan:2018zdc}. $\mu^2$ is not energy-dependent, which is reflected in the tunes for 
$\sqrt{s}=7,13\,\mathrm{TeV}$. For high centre-of-mass energies, the $\chi^2$ value also is
less sensitive to the input from various combinations of Monte Carlo runs used for tuning. 
For smaller centre-of-mass energies, we note an increased sensitivity to the chosen runs. 
In a similar approach as\,\cite{Gieseke:2012ft} we fix $\mu^2$ to the tightly 
constrained tuned values for 7\,TeV and 13\,TeV. Both favour small ranges of $\mu^2= 1.1$ 
but with different values for $p_{\perp}^{\mathrm{min}}$ independent of the run combinations 
used for tuning. With $\mu^2$ fixed we then re-tune $p_{\perp}^{\mathrm{min}}$ to the 
different centre-of-masss energies.  
The median values for $p_{\perp}^{\mathrm{min}}$ and the corresponding spread of the values and the 
ranges for $\mu^2$ resulting from the different run combinations are listed in 
Tab.\,~\ref{table:pTmin}. We note here that these values do not correspond to the best tune values which will result in the overall minimal value for $\chi^2$. 
Important at this stage is to find a sensible parameterization which approximately captures 
the energy dependence of the $p_{\perp}^{\mathrm{min}}(s)$ parameter at different centre-of-mass energies.

\begin{table}
\centering
\begin{tabular}{c||c|c|c|c|c|c}
 $\sqrt{s}\,\mathrm{[GeV]}$ & 200  & 500 & 900 & 1800 & 7000 & 13000  \\
\hline
\hline
 $p_{\perp}^{\mathrm{min}}\,\mathrm{[GeV]}$ & 1.80 & 1.6 & 1.75 & 1,95 & 2.91 & 3.5 \\
\hline
 $\mu^2\,\mathrm{[1/GeV^2]}$ & 1.0-2.2 & 1.1-2.5 & 1.0-2.0 & 1.0-1.6 & 1.05-1.15 & 1.0-1.2 
\end{tabular}
\caption{Tune values for $p_{\perp}^{\mathrm{min}}$ and the ranges of $\mu^2$.}
\label{table:pTmin}
\end{table}

\begin{figure}
\includegraphics[width=0.48\textwidth]{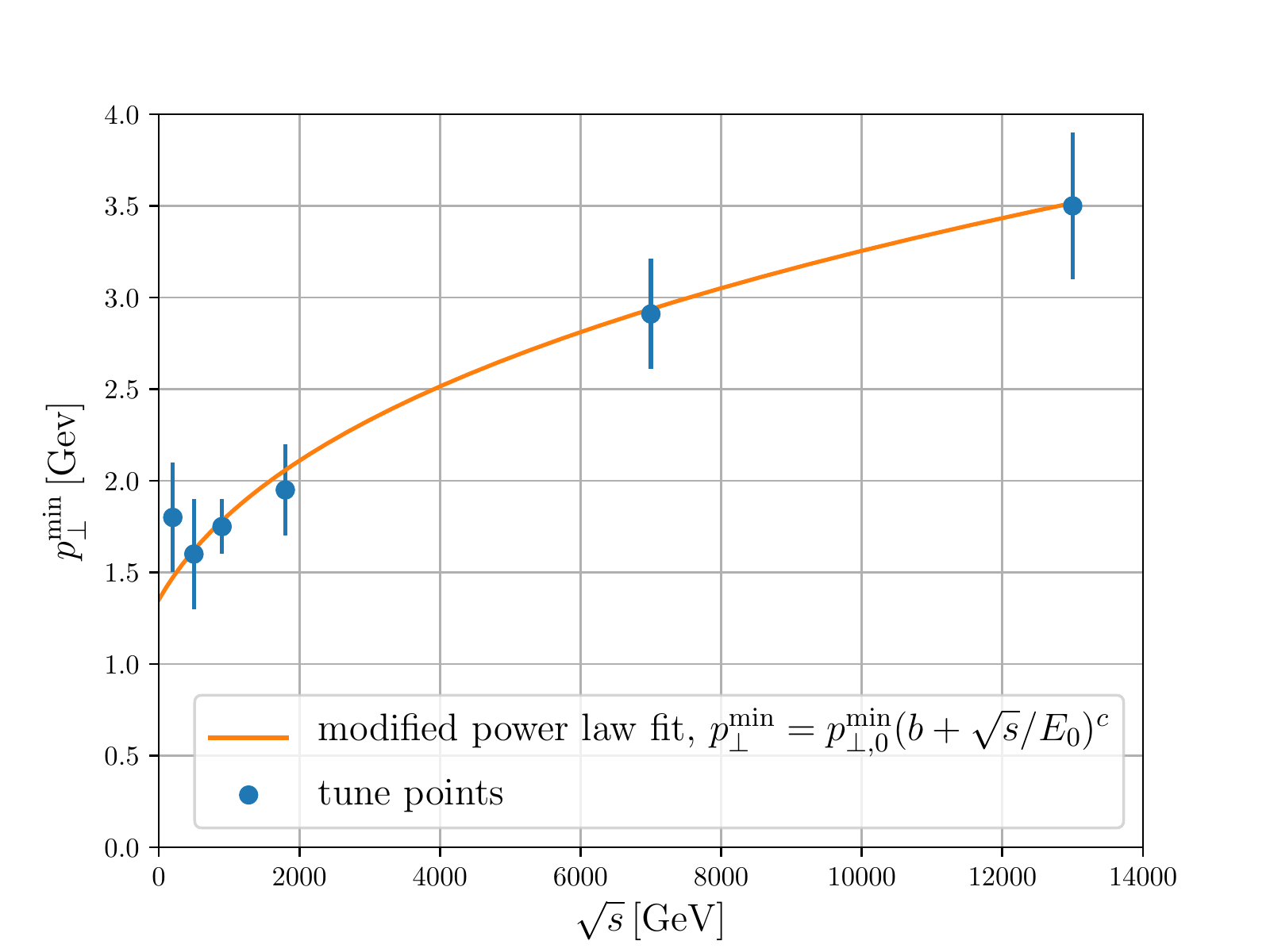}
\caption{Energy extrapolation of $p_{\perp}^{\mathrm{min}}$. The blue dots are determined from dedicated tunes to data at the given energy. Here, the possible ranges of the $p_{\perp}^{\mathrm{min}}$ value are displayed with error bars. }
\label{fig:pTmin}
\end{figure}

In order to account for a good description of MB and UE data over the considered energy 
range between 200\,GeV and 13\,TeV it is necessary to modify the existing power law 
parametrization of the $p_{\perp}^{\mathrm{min}}$ parameter~\cite{Gieseke:2012ft} 
and introduce an energy offset to account for the $p_{\perp}^{\mathrm{min}}$ 
values at $\sqrt{s}<900\,\mathrm{GeV}$. The modified power law is given in Eq.\,~\ref{eq:powerlaw}. A plot with the tuned $p_{\perp}^{\mathrm{min}}$ 
values and the resulting fit are shown in Fig.\,~\ref{fig:pTmin}. 
We find that the $p_{\perp}^{\mathrm{min}}$ points can be fitted with the 
parameter values summarized in Tab.\,\ref{table:fit} if we demand that the parametrization
must reproduce the $p_{\perp}^{\mathrm{min}}$(s) values for $\sqrt{s}=7,13$\,TeV opting 
for an improved description at higher centre-of-mass energies taking into account a less 
optimal fit of $p_{\perp}^{\mathrm{min}}(s)$ at $\sqrt{s}=200$\,GeV.

\begin{table}
\centering
\begin{tabular}{c|c|c|c}
   $p_{\perp,0}^{\mathrm{min}}\,\mathrm{[GeV]}$ & $b\,\mathrm{[GeV]}$ & $c$ & $E_0\,\mathrm{[GeV]}$  \\
\hline
\hline
   2.87 & 622 & 0.31 & 7000 
\end{tabular}
\caption{Parameter values used in the modified energy extrapolation of the $p_{\perp}^{\mathrm{min}}$ parameter.}
\label{table:fit}
\end{table}
With the modified parametrization, the MPI model of Herwig works at all centre-of-mass 
energies considered and the reoccurring issue that the hard cross-section exceeds 
the total cross-section is no longer present for small $\sqrt{s}$.

The remaining parameters of the MPI model 
($N_{\mathrm{ladder}}$, $B_{\mathrm{ladder}}$, 
$R_{\mathrm{Diffraction}}$) and the two colour reconnection probabilities 
($p_{\mathrm{Reco}}$, $p_{\mathrm{RecoBaryonic}}$) are tuned to MB and UE data measured 
at the LHC at 7\,TeV, and 13\,TeV. We restrain from tuning
to single observables and instead focus on an overall reduction of the total
$\chi^2$ measure.
The resulting set of parameters which were tuned with the new parametrization for
the $p_{\perp}^{\mathrm{min}}$-values are listed in Tab.\,\ref{table:parameters1}.

In addition to the energy-independent tune, we provide individual tunes for the full set of MPI parameters for 7\,TeV and 13\, TeV which were performed with the Autotunes framework for tuning\,\cite{Bellm:2019owc}. 
The resulting parameter values are also listed in Tab.\,\ref{table:parameters2}.

All performed parameter scans showed no dependence of the goodness of fit
value on the $B_{\mathrm{ladder}}$ parameter which we set to zero. 
The tuned value for the ladder multiplicity varies between 0.6 and 0.699 which can be seen
as a sign for weak energy dependence and we conclude that Eq.\,\ref{eq:ladderMult}
captures the energy dependence of the ladder multiplicity.
The value of the diffractive ratio is mainly driven by the $\mathrm{d}\sigma/\mathrm{d}\Delta\eta^F$  observables from \cite{Aad:2012pw,Chatrchyan:2014fsa} and the 
diffractive cross-sections measurements, $\sigma_{\mathrm{SD}}$, $\sigma_{\mathrm{DD}}$ 
from \cite{Abelev:2012sea} at 7\,TeV. 
More data at different $\sqrt{s}$ is needed to correctly assess the energy 
dependence of the diffractive cross-section\,(see e.g. \cite{Rasmussen:2018dgo}). 
The high value for the reconnection probabilities is due to the requirement of the 
colour reconnection model that a certain rapidity configuration of quarks within clusters
has to be met in order to be considered for reconnection. The effective 
reconnection probability is therefore lower. The probability for baryonic reconnection
is mainly driven by flavour observables at 7\,TeV. As can be seen in Tab.\,\ref{table:parameters2}, $p_{\mathrm{RecoBaryonic}}$
drops significantly from 7\,TeV to 13\,TeV. Nonetheless, it is possible to have more baryonic clusters due to
the increased multiplicity at 13\,TeV.

In Sec.~\ref{sec:results} we show some exemplary plots where we compare 
Herwig 7.1.5 to the modified MPI model. 
The parameters presented in this paper are set as the default in the new release of 
Herwig 7.2 and can be used as a well-motivated baseline describing general properties of MB and UE data over a wide range of centre-of-mass energies. 

The modified power-law parametrization of the energy dependence of the
 $p_{\perp}^{\mathrm{min}}$ parameter with the given values is used as the default
in the new Herwig release 7.2.

\section{Results}
\label{sec:results}
In this paper, we have presented various changes to the MPI model in Herwig. With the mentioned
modifications in Sec.\,~\ref{sec:modifications} we see a good description of MB and UE observables combined across all considered centre-of-mass energies. In this section, we discuss the resulting influence of the modifications described in Sec.\,~\ref{sec:modifications} on some handpicked observables. A complete set of all observables
with the current model will be made available on the webpage\,\cite{webpage}.

\begin{figure}[t]
\includegraphics[width=0.48\textwidth]{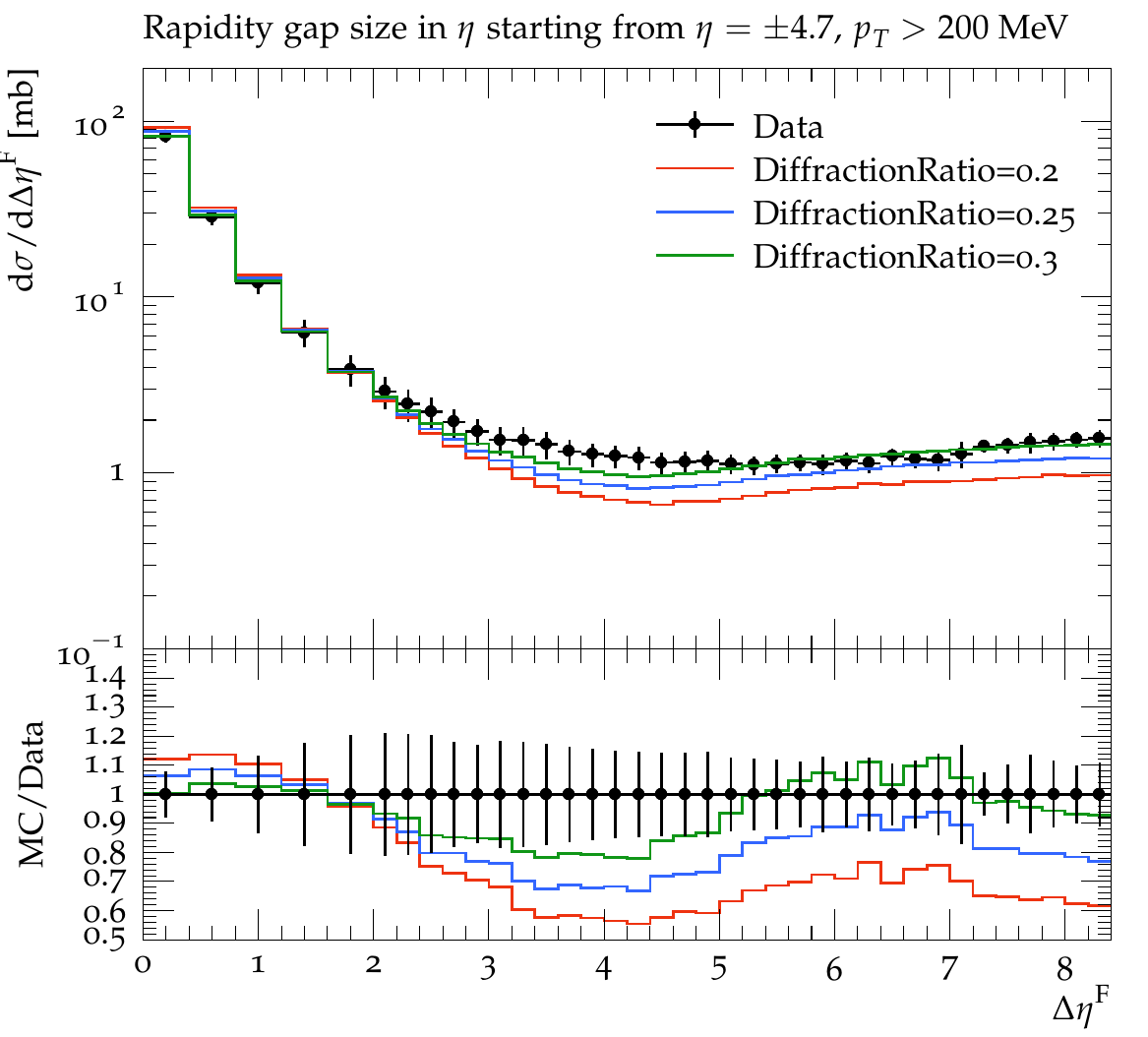}
\caption{Impact of the DiffractionRatio parameter choices in comparision to differential cross-section data measured in \cite{Khachatryan:2015gka}.}
\label{fig:diffratio}
\end{figure}
\textbf{DiffractionRatio:}
The parameter gives an easier handle on the fraction of diffractive events and can be tuned to
observables sensitive to diffractive events like the rapidity gap cross-section 
$\mathrm{d}\sigma/\mathrm{d}\Delta\eta^F$ as measured by CMS and ATLAS at 7\,TeV. 
In Fig.\,\ref{fig:diffratio} we show the $\mathrm{d}\sigma/\mathrm{d}\Delta\eta^F$ observable for 
 three different values of the DiffractionRatio parameter.
We see that varying the DiffractionRatio modifies the differential cross-section with respect
to events containing large rapidity gaps $\Delta\eta^F$ but leaves the region with 
$\Delta\eta^F<2$ largely invariant which is dominated by non-diffractive particle production.

\begin{figure}[t]
\includegraphics[width=0.48\textwidth]{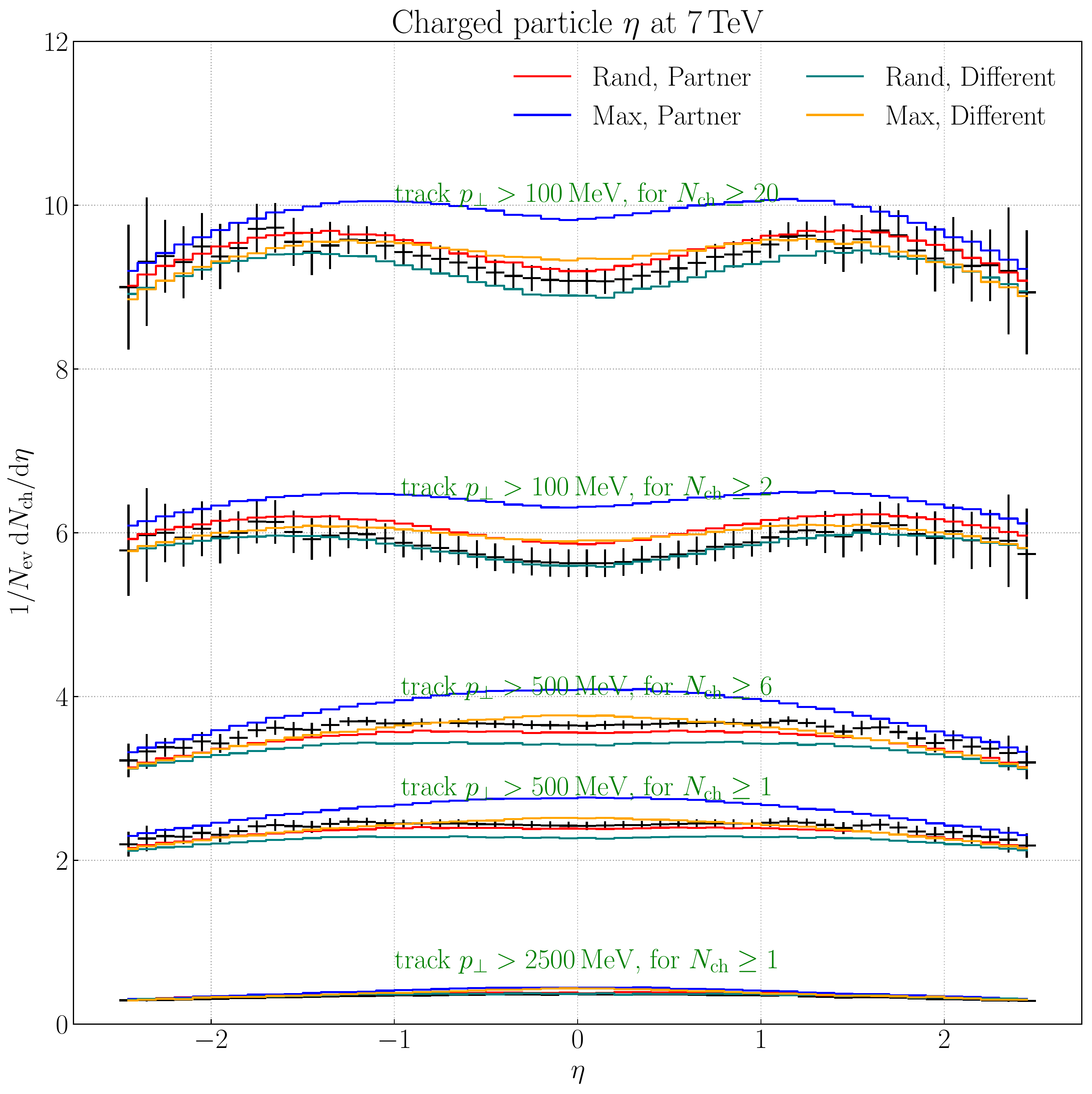}
\caption{The charged-particle multiplicity is plotted against the rapidity for multiple cuts on the hardest track transverse momentum and number of charged particles. Data is taken from \cite{Aad:2010ac}. This observable is sensitive to the choices that are employed as the starting conditions of the parton shower process. The four choices are described in Sec. \ref{sec:modifications}. Height differences are easily modified in the tuning process, but shape differences prefer choices with a random colour partner for gluons in the hard process.  \label{fig:partnerfinder}}
\end{figure}
\textbf{Partner and Scale Choice:}
The impact of the partner and scale choice is shown in Fig.~\ref{fig:partnerfinder}. In comparison to data from \cite{Aad:2010ac} showing the charged multiplicity as a function of rapidity in MB events, we give results from the four choices implemented in Herwig. We note, that in the tuning process the height can be modified by multiple variables, like the value of the strong coupling or the ladder multiplicity. 
The shape of the distributions, however, depends more on the underlying dynamics and in this case the shower evolution. The best reproduction of the central plateau at $p_{\perp}\geq500$\,MeV leading tracks and
 the dip for low energy leading tracks can be achieved by choosing the evolution 
 partner of the gluon randomly between one of the two colour connected partners.

\begin{figure}
\includegraphics[width=0.48\textwidth]{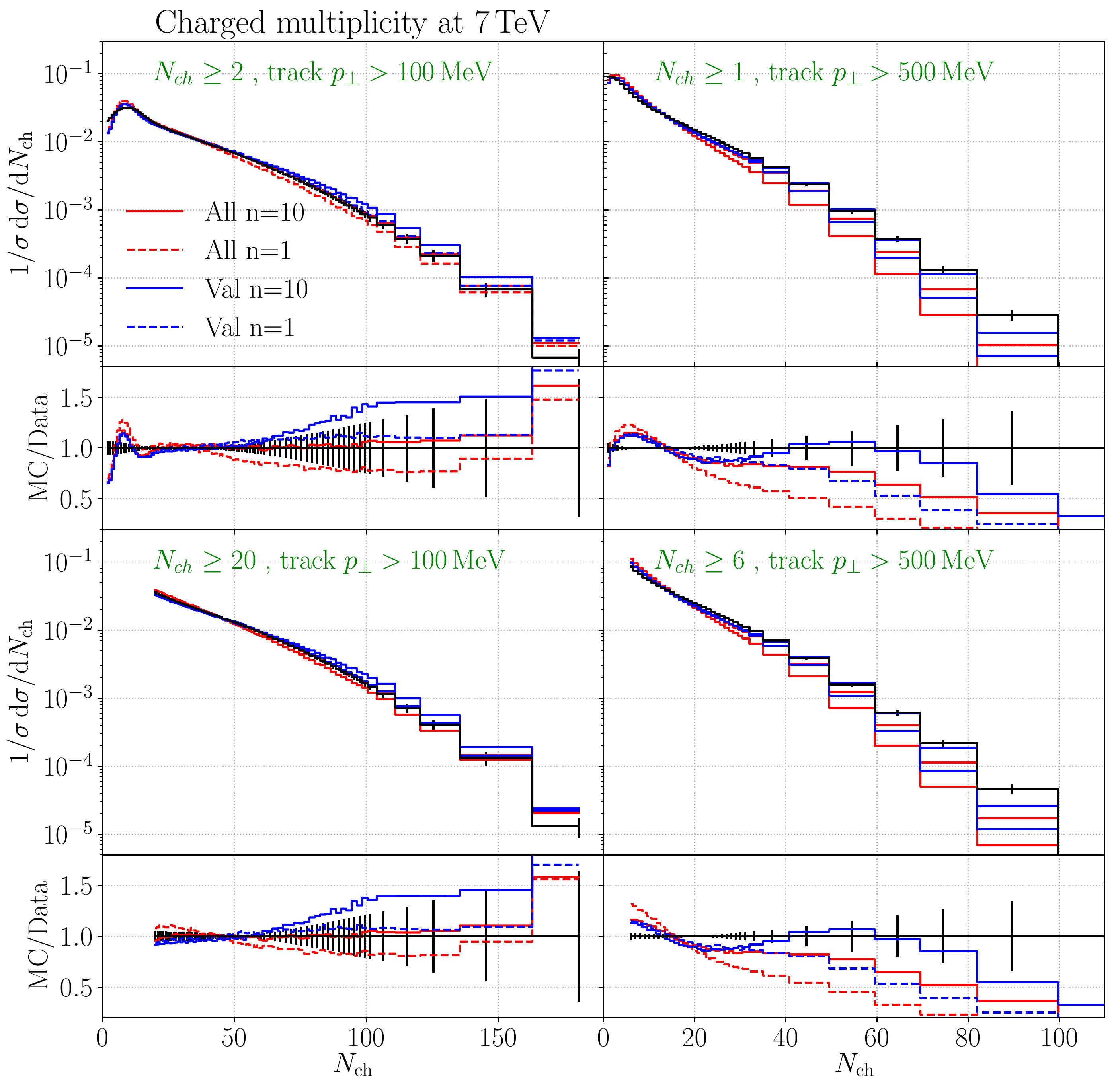}
\caption{We show the model comparison to normalized cross-section data from 
\cite{Aad:2010ac} as a function of the number of charged particles in the event. 
The figure depicts the influence of the possibility of 
extracting only valence quarks (blue) or all light flavours (red) at the stage of the 'dummy' process. 
It further shows the possibility of multiple trials $n$ in the generation of additional scatters.  \label{fig:valence}}
\end{figure}
\textbf{Dummy Process using Valence Quarks:}
Fig.~\ref{fig:valence} shows the impact of choosing all light quark flavours to be initial states to the dummy process starting the generation of soft and/or hard additional scatters. Further, we allow each additional scatter to be reproduced $n$ times if the showering of the additional hard process ends up in a momentum configuration that is not supported by the current energy fractions extracted from either of the protons. It is visible that depending on the choice a change of up to 50\% in the tails of the charged particle multiplicities can be generated. As can be expected, 
allowing multiple tries to regenerate configurations contributes to higher particle multiplicities. 
Further, the restriction to "only valence" quark extraction and the reduction of forced splittings also contributed to an increased charged particle multiplicity as on average with higher energy additional scatters are possible. 
This figure also illustrates the possible variation one should expect from such subtle changes.

\begin{figure}
\includegraphics[width=0.48\textwidth]{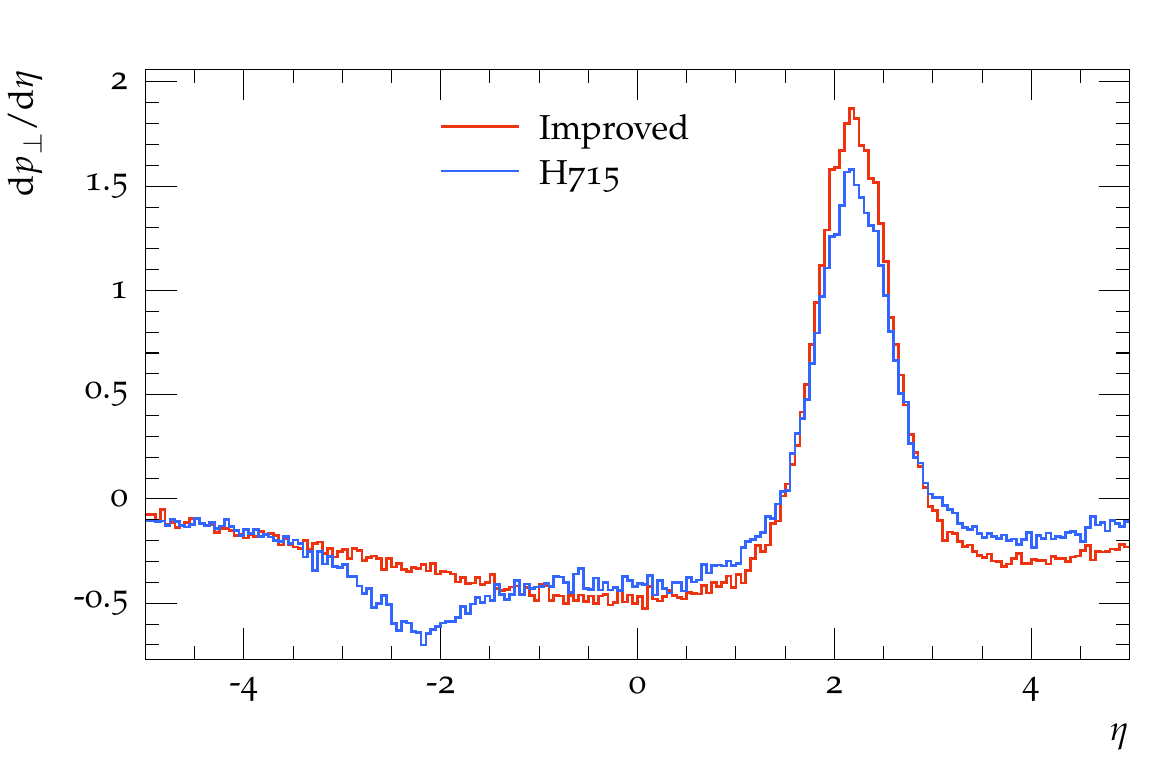}
\caption{Comparison between Herwig 7.1.5 and the improved version of the model. Plot of the minijet correlations from \cite{Azarkin:2018cmr}.
}
\label{fig:minijets}
\end{figure}

\textbf{Kinematics of Soft Ladder:}
The plot from \cite{Azarkin:2018cmr} with the
new kinematic description is shown in Fig.\,~\ref{fig:minijets}.
The new kinematics is necessary to account for more fluctuations in the rapidity values for
the produced particles in the soft ladder and removes the unphysical anticorrelation seen in\,\cite{Azarkin:2018cmr} for $\eta \approx -2$. The change to the kinematic description does not necessarily result in a better description of data but was necessary for a
physically more sound model.

\begin{figure}[t]
\includegraphics[width=0.48\textwidth]{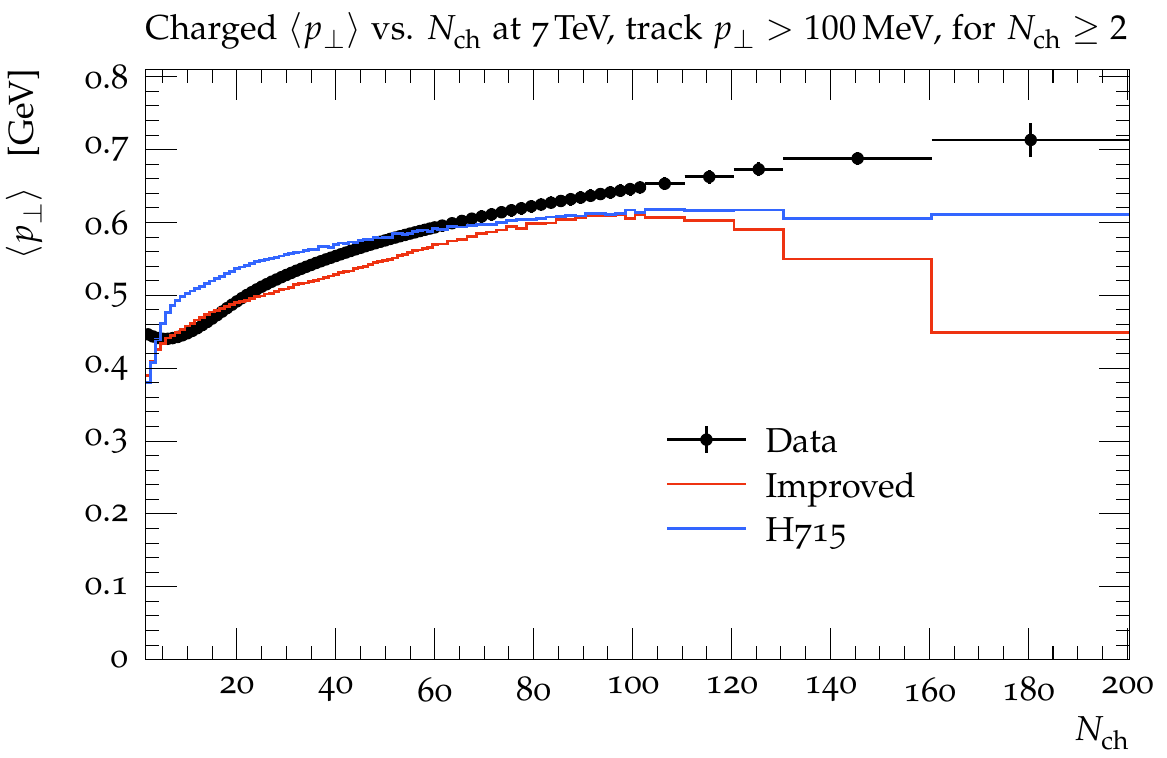}
\caption{Comparison between the old and the improved version of the model. The plot shows the average transverse momentum $\langle p_{\perp}\rangle$ against the charged particle multiplicity $N_{\mathrm{ch}}$ observable as measured by ATLAS \cite{Aad:2010ac}.
}
\label{fig:avgPtNch}
\end{figure}
\textbf{Soft Ladder Transverse Momentum:}
The change in the sampling of the transverse momentum of the particles produced during soft interactions is best shown with the $\langle p_{\perp}\rangle$ vs. $N_{\mathrm{ch}}$ observable.
although the description seems to worsen in the high $N_{\mathrm{ch}}$ tail the majority of the events has multiplicities of up to 100 charged particles. With the new assignment of $p_{\perp}$ the 
\textit{bump} for $N_{\mathrm{ch}}<40$, which was present in the old version disappears since we notably shift the 
$p_{\perp}$ of the produced particles towards lower values which is in alignment with the measurements in that multiplicity region. The resulting plot is shown in 
Fig.\,~\ref{fig:avgPtNch}. For the $N_{\mathrm{ch}}>100$ region, we produce too many soft particles. 
A detailed study concentrating on high multiplicity events would be needed to remedy that problem.


\section{Conclusion and Outlook}
\label{sec:conclusion}
In this paper, we have summarized various changes to the MPI model in Herwig.
We have introduced a parameter called DiffractiveRatio which gives an easy handle on the fraction of diffractive events in a collision. We have reviewed and changed the model for soft interactions with a well-motivated change to the $p_{\perp}$ distribution
of the partons produced in a soft ladder. To achieve a better description of data we changed the partner and scale setting for gluons in the starting conditions of the angular ordered parton shower. 
Furthermore, we tuned the modified model to MB and UE data covering an energy range between 200\,GeV and 13\,TeV which led to a modification of the existing energy
extrapolation of the MPI model. The resulting tune gives generally a good description of all observables with only one set of parameters, albeit due to the nature of the combined tune, there are some aspects which are not described satisfactorily.

Especially in the high multiplicity region, we fail to describe the data correctly for observables like $\langle p_{\perp}\rangle$ vs. $N_{\mathrm{ch}}$. 

With the rising number of differential measurements at the LHC, especially flavour observables
and measurements probing the high multiplicity region of hadronic collisions, it becomes
harder for multi-purpose Monte Carlo event generators to capture all phenomena
observed in the measurements and to account for a sensible description of data. In this paper,
we have taken the approach to tune the free parameters of the MPI model to as many observables
as possible where the only guidance was the overall $\chi^2$ value. 

The modifications described in this paper are a step towards a better understanding of soft and non-perturbative physics at high energy colliders. Next steps are needed. 
The inclusion of the diffractive cross-section in the eikonal model has been improved but a full eikonalisation and the implications to the generation details of then possible multiple diffractive processes has not been addressed in this work. 
Furthermore, to correctly assess the energy dependence of the diffractive cross-section, more data at
13\,TeV is needed.

The dummy process to start the generation of multiple scatters 
is a non-necessary construction that perhaps needs to be overcome at some point.
Also the full connection with the work done in \cite{Bellm:2019wrh} and 
a rethinking of the colour connections of the soft ladders according to 
new measures, i.e. minimal masses,  may be introduced.

\appendix
\section{Parameter values}
In this appendix, we present the tuned parameter sets discussed in Sec. \ref{sec:tuning}.
Note that the parameter set of Tab.\ref{table:parameters1} will become the default in the next Herwig release 7.2 and the parameters in Tab. \ref{table:parameters2} are dedicated 
tunes to 7\,TeV and 13\,TeV LHC data.
\begin{table}[H]
\centering
\begin{tabular}{c|c}
  Parameter & Value  \\
\hline
\hline
$N_{\mathrm{ladder}}$  &  0.6838  \\
 $R_{\mathrm{Diffraction}}$ & 0.187 \\
 $p_{\mathrm{Reco}}$ & 0.970   \\
 $p_{\mathrm{RecoBaryonic}}$ & 0.626  \\
 $\mu^2$ & 1.1 \\
$p_{\perp,0}^{\mathrm{min}}$ & $2.82$ \\
$b$ & 622.203\\
$c$ & 0.31 \\
$E_0$ & 7000\\
\end{tabular}
\caption{Parameters of the energy independent UE and MB tune with the parametrization of $p_{\perp}^{\mathrm{min}}$ from Eq.\ref{eq:powerlaw}.}
\label{table:parameters1}
\end{table}

\begin{table}[H]
\centering
\begin{tabular}{c|c|c}
  Parameter & 7\,TeV & 13\,TeV \\
\hline
\hline
$N_{\mathrm{ladder}}$ & 0.601 & 0.699\\ 
 $R_{\mathrm{Diffraction}}$ & 0.2574 & 0.167\\
 $p_{\mathrm{Reco}}$ & 0.99 & 0.811\\
 $p_{\mathrm{RecoBaryonic}}$ & 0.898 & 0.496\\ 
 $\mu^2$ & 1.1 & 1.1 \\ 
$p_{\perp}^{\mathrm{min}}$ & 2.827 & 3.97\\ 
\end{tabular}
\caption{Energy dependent parameter values of the UE and MB tune for 7\,TeV and 13\,TeV.}
\label{table:parameters2}
\end{table}

\section*{Acknowledgements} 

We would like to thank the Herwig collaboration and especially Cody B. Duncan, Miroslav Myska, Andrzej Siódmok and Peter Richardson for valuable discussions on the topic. We thank Steffen Gay for his contributions in the early stages of this work.
This work was supported by the MCnetITN3 H2020 Marie Curie Initial Training Network, grant agreement 722104. This project has also received funding from the European Research Council (ERC) under the European Union’s Horizon 2020 research and innovation programme, grant agreement No 668679. This work has been supported by the BMBF under grant number 05H18VKCC1.

\bibliography{journal}

\end{document}